# Poly(ionic liquid)-Derived Carbon with Site-Specific N-Doping and Biphasic Heterojunction for Enhanced CO$_2$ Capture and Sensing

Jiang Gong, Markus Antonietti and Jiayin Yuan*

**Abstract:** CO$_2$ capture is a pressing global environmental issue that drives scientists to develop creative strategies for tackling this challenge. The concept in this contribution is to produce site-specific nitrogen-doping in microporous carbon fibers. It creates a carbon/carbon heterojunction by using poly(ionic liquid) (PIL) as "soft" activation agent that deposits nitrogen species exclusively on the skin of commercial microporous carbon fibers. Such carbon-based biphasic heterojunction amplifies the interaction between carbon fiber and CO$_2$ molecule for unusually high CO$_2$ uptake and resistive sensing.

Growing concerns on climate changes have urged researchers worldwide to develop effective methods for environmental remediation and monitoring[1]. Processes involving CO$_2$ as the predominant greenhouse gas are considered to be most pressing[2]. Conventional technologies for CO$_2$ removal by using bases involve aggressive chemical media[3]. As an alternative, various CO$_2$ sorbents have been studied over the past two decades, such as metal-organic frameworks (MOFs)[4], covalent organic frameworks[5], zeolite[6], carbon[7] and more[3c,8]. Once collected, CO$_2$ can be sequestered, and there is also current interest in catalytic conversion of CO$_2$ into industrial raw materials[9]. Despite all the accomplishments, the pursuit of efficient and cost-effective CO$_2$ sorbents remains a critical issue. Equally important, simple CO$_2$ detection techniques preferentially at ambient conditions are favorable in many cases, *e.g.*, indoor air-quality control and engine exhaust monitoring[10]. Common CO$_2$ sensing technologies[11] include electrochemical measurement, nondispersive infrared sensing, *etc*. These techniques have associated limitations, *e.g.*, the requirement of sophisticated instruments or pre-sensing gas treatment. From a viewpoint of materials choice and device design, materials that are not only useful in CO$_2$ capture but also inherently effective in sensing are rare and a potential game changer.

Nitrogen-doped porous carbons have recently been explored with great interests in environmental and energy applications due to their tunable physicochemical properties[12]. Nitrogen dopant in a carbon network alters the electronic band structure thus the conductivity and oxidative resistance[13]. CO$_2$ capture over carbon is reported to depend on the surface chemical state and porosity, and logically the incorporation of nitrogen into carbon enhances the affinity between acidic CO$_2$ molecule and basic nitrogen-rich surface[14]. N-doped carbons with large specific surface area are therefore well examined and constitute one of the few golden standards to beat[15]. Their routine synthesis involves chemical/physical activation of preformed N-doped carbon, carbonization of N-rich precursor in the presence of template, or thermal doping of porous carbon by reactive N-rich gas. Besides, elaborating the relationship between surface-exposed nitrogen active sites and sorption/monitoring property of carbon is crucial to identify the function of nitrogen active sites and improve CO$_2$ capture/sensor performance. Recently the heterojunction between N-doped carbon and ordinary all-carbon nanostructure, *e.g.*, carbon nanotube (CNT), is found to create a spontaneous electron flux from carbon to N-doped carbon[16]. This charge separation results in a stable dispersion of colloidal carbon superstructure in aqueous solution.

To explore the carbon/carbon heterojunction and transferred charge for sorption application, herein we synthesize nitrogen-doped microporous carbon fibers (NPCFs), where nitrogen is merely deposited on the surface of microporous carbon fibers (PCFs) in one step using PIL as coating/mild porosion agent. Due to the as-formed core/sheath structure, these NPCFs in spite of an overall low nitrogen content (< 0.5 wt %) display exceedingly high CO$_2$ uptake and detection sensitivity, thereby for the first time revealing a new CO$_2$ capture mechanism enabled by previously underestimated electronic effects[17].

Figure 1a schematically illustrates the synthetic approach to NPCFs using a cationic PIL (Figures S1–S3) as a multifunctional modification agent. The choice of this PIL is of key importance to tailor properties of the final carbon material. Firstly, PILs are well-known for their strong interaction with substrate to form a stable thin coating that secures homogeneous surface modification of complex objects[18]. Secondly, the aromatic imidazolium cation contains abundant nitrogen atoms that favorably enter the carbon product *via* rearomatization. Thirdly, the chosen PIL itself under carbonization condition generates a highly microporous carbon due to a self-templating mechanism involving bis(trifluoromethane sulfonyl)imide (Tf$_2$N) anion[19]. These three aspects in a single reagent favor the formation of a thin N-doped microporous carbon layer exclusively on PCF surface.

Briefly, PCFs and the PIL were mixed in *N,N*-dimethylformamide (DMF). Owing to the cation-π interaction, PIL chains adsorb onto PCF surface to firstly build up a monolayer coating. The relatively large PIL chains are size-excluded from the bulk micropores of PCF and occupy the PCF surface only. The initial monolayer coating improves the polarity of PCF surface and makes it better wetted by DMF. The formed carbon/polymer core/sheath composite (Figure S4) is termed as PCF@PIL-*x* (*x* = 5, 10 and 20), where *x* denotes the PIL/PCF mass ratio (%) when mixed in DMF. Next, the dry composite was calcined at 750 °C to convert the PIL into N-doped microporous carbon. The core/sheath hybrid PCF@N-doped carbon is named as NPCF-*x*, while the carbon powder prepared from merely PIL is denoted as C-PIL. The carbonization yield of

[Dr. J. Gong, Prof. M. Antonietti, Prof. J. Yuan
Department of Colloid Chemistry
Max Planck Institute of Colloids and Interfaces
14476 Potsdam, Germany

Prof. J. Yuan
Department of Chemistry and Biomolecular Science, & Center for Advanced Materials Processing
Clarkson University
13699 Potsdam, USA
E-mail: jyuan@clarkson.edu]

Supporting information for this article is given via a link at the end of the document.

the native PIL is 21.7 wt %. In the PCF@PIL-*x* composite, PIL exists only as a thin coating layer of < 1.5 wt % of PCF; thereby the carbonization of composite yields no detectable mass loss (< 5%). The presence of nitrogen is confirmed by elemental analysis to be 0.21, 0.32 and 0.42 wt % for NPCF-5, -10 and -20, respectively (Table S1), thus excluding the direct composition influence of nitrogen content on $CO_2$ sorption behavior discussed later.

Scanning electron microscopy (SEM) visualizes the structural evolution along the synthetic route. Microscopically, the fibers of ca. 25 μm in diameter and several centimeters in length appear identical at different stages (Figures 1b,1c,S5), indicating that the fibril morphology is favorably preserved. This morphology preservation is important, since fibers are more easily integrated into functional devices than powders (Figure S6). On the nanoscale, structural variations are observed. NPCF-10 as a representative example presents an evenly distributed, nanoporous texture on surface, while PCF bears a comparatively closed surface (Figures 1a left,S7). A closer front view on NPCF-10 reveals that the surface is patched with stripes of 15–30 nm in width separated by numerous nanocanyons (Figure 1d). These patches are already observed in PIL-coated PCF (Figures 1a middle,S8), that is, they are a feature of coating process. Energy-dispersive X-ray (EDX) maps on the fiber surface (Figures 1e–1g) and along cross-section (Figure S9) demonstrate that nitrogen distribution is uniform and exclusively on surface. SEM cross-sectional view (Figures 1h–1j) shows the N-doped porous carbon nanolayer on fiber surface that appears different from the bulk. Similar core/sheath structure is observed in NPCF-5 and -20 (Figure S10).

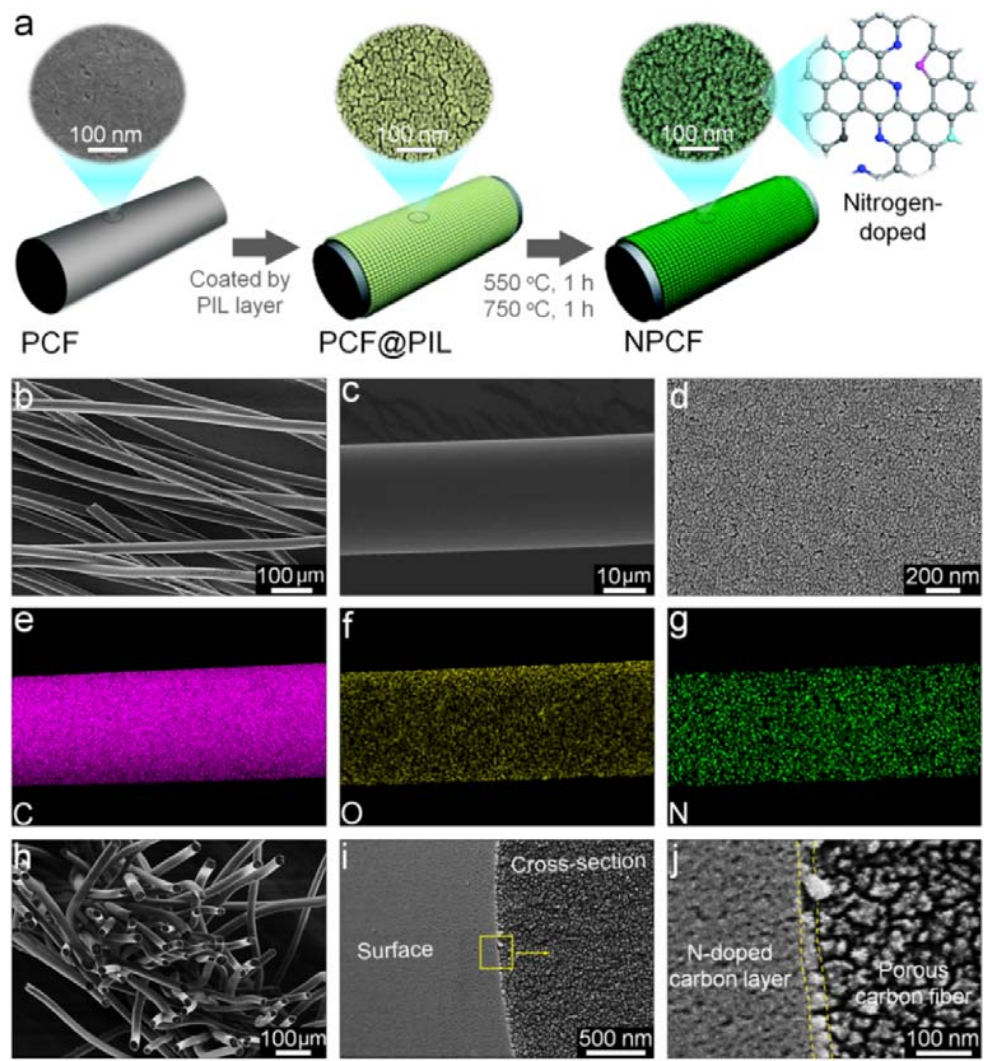

**Figure 1** (a) Schematic illustration of NPCF preparation. SEM images (b–d for front view and h–j for cross-sectional view) and EDX maps (e–g) of NPCF-10.

$N_2$ adsorption/desorption measurements were conducted to analyze textural properties (Figures 2a,2b,S11, Table S2). C-PIL exhibits a combined type I/IV physisorption isotherm with a high sorption capacity at a low relative pressure ($P/P_0$), revealing the presence of abundant micropores and characterized by a specific surface area (*S*) of 707.6 m$^2$ g$^{-1}$ and a pore volume (*V*) of 0.382 cm$^3$ g$^{-1}$. The detectable type-H4 hysteresis loop at $P/P_0$ = 0.4–0.8 is owing to the filling and emptying of mesopores. In comparison, PCF presents a type I physisorption isotherm, and the *S* and *V* are 800.8 m$^2$ g$^{-1}$ and 0.396 cm$^3$ g$^{-1}$, respectively.

Likewise, NPCFs deliver a type I physisorption isotherm, and the pore size distributions ranging from 0.6 to 2.2 nm affirm the dominant presence of micropores. Remarkably, the $S$ and $V$ increase to 1024.2 m$^2$ g$^{-1}$ and 0.474 cm$^3$ g$^{-1}$ for NPCF-5, 1476.3 m$^2$ g$^{-1}$ and 0.583 cm$^3$ g$^{-1}$ for NPCF-10, and 1319.0 m$^2$ g$^{-1}$ and 0.523 cm$^3$ g$^{-1}$ for NPCF-20. There is thus an additional pore opening effect of N-doped carbon layer on PCF. This is attributed to pore formation in skin layers: the surface of PCF is dense (Figure 1a left); PILs act as mild activation agents to promote pore formation in carbons at elevated temperature[20], since their decomposition fragments open and activate the substrate layer (Figure 2c). As an indirect proof, the physical mixture of PCF and C-PIL only has an $S$ of 802.0 m$^2$ g$^{-1}$ (Figure S12). With these performance values, NPCFs are superior to N-doped CNT[21] and graphene[22] that display similar core/sheath structure but notably lower $S$ (< 250 m$^2$ g$^{-1}$).

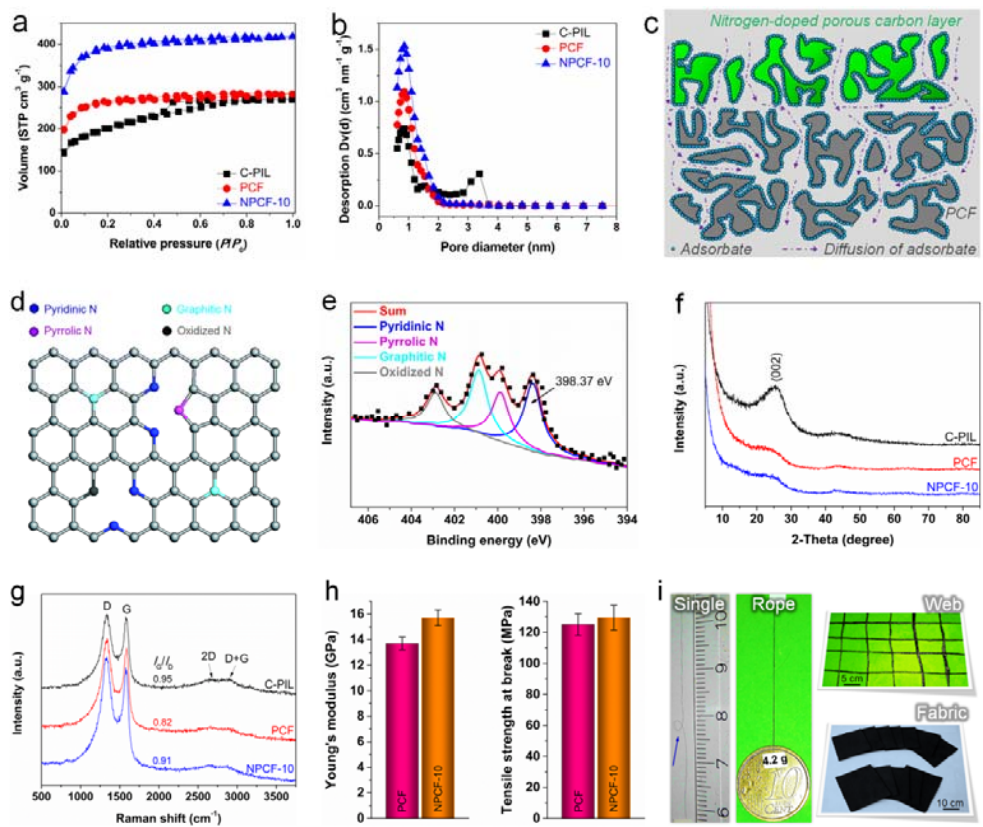

**Figure 2** (a) N$_2$ adsorption/desorption isotherms at 77 K and (b) the corresponding pore size distribution plots. (c) Scheme of NPCF-10 cross-sectional structure by depositing N-doped porous carbon nanolayer on PCF surface. (d) Scheme of nitrogen species. (e) High-resolution N 1s XPS curve of NPCF-10. (f) XRD patterns. (g) Raman spectra. (h) Comparisons of Young's modulus and tensile strength of NPCF-10 with PCF. (i) Photographs of a flexible single fiber with a loop (pointed by a blue arrow), a robust rope built up from six fibers, easy-to-handle webs and fabrics.

X-ray photoelectron spectroscopy (XPS) spectra are presented in Figure S13. High-resolution N 1s XPS peak is deconvoluted into four components[23]: pyridinic, pyrrolic, graphitic and oxidized N (Figure 2d). Pyridinic N (31.6–42.7%) and graphitic N (26.4–39.2%) are the dominant types (Figure 2e, Table S3). X-ray diffraction (XRD) was employed to investigate the phase structure (Figure 2f). The appearance of two weak diffraction peaks centered at 26° and 42° discloses a poorly stacked graphitic structure. Raman spectroscopy was applied to complement our insights into phase structure (Figure 2g). The G band at ca. 1575 cm$^{-1}$ and D band at ca. 1340 cm$^{-1}$ correspond to the ordered $sp^2$ carbon structure and defective structure, respectively. A low ratio of $I_G/I_D$ (0.91) implies a high disorder of NPCF-10. It is worth noting that the carbonization conditions (*e.g.*, temperature and time) influence the physical and chemical nature of NPCF-10 only to a certain extent (Table S4).

Additionally, NPCF-10 exhibits several favorable secondary properties such as high electrical conductivity, good mechanical properties and excellent handling. The conductivity is ~1300 S m$^{-1}$, 12 times that of N-doped CNT[21]. In spite of a high porosity, its tensile strength at break is up to 129.5 MPa (Fig. 2h), higher than that of graphene[24] or CNT[25] fiber, while its Young's modulus (15.7 GPa) is at least three orders of magnitude greater than that of CNT fiber[25]. The mechanical flexibility allows NPCF-10 to bend into a loop without any damage or to process into well-defined shapes, *e.g.*, a robust rope and easy-to-handle webs and fabrics (Figure 2i). In this regard, NPCF-10 prevails over traditional powder carbon and allows simpler access to advanced applications[26]. Besides, the employment of commercial PCF (~30 euro kg$^{-1}$) promises scalable transfer.

Figure 3a depicts the CO$_2$ adsorption isotherms at 273 K. NPCF-10 shows an unprecedentedly high adsorption capacity of 6.9 mmol g$^{-1}$, meaning 1.0 g of NPCF-10 adsorbs 0.3 g of CO$_2$. This is substantially larger than its two structural components,

PCF (2.3 mmol g$^{-1}$) and C-PIL (2.8 mmol g$^{-1}$), indicating the synergistic effect of the dyadic coupling of two carbons. $CO_2$ adsorption capacities of NPCF-5 and -20 (Figure 3b) are 4.1 and 5.4 mmol g$^{-1}$, respectively, lower than that of NPCF-10. As NPCF-20 has close textural parameters (Table S2), this is a hint that the high $CO_2$ uptake is not primarily related to pore structure or nitrogen dopant that anyway provides only a few extra binding sites. Under similar condition, $N_2$ adsorption capacity of NPCF-10 reaches 0.49 and 0.38 mmol g$^{-1}$ at 273 and 298 K, respectively (Figure 3c), considerably lower than that for $CO_2$. The initial slopes of $CO_2$ and $N_2$ adsorption isotherms are used to estimate the selectivity. The apparent $CO_2/N_2$ selectivity at 273 and 298 K is 44 and 17, respectively. We compared the $CO_2$ uptake and $CO_2/N_2$ selectivity of NPCF-10 with previous carbon[15,27], MOFs[28] and porous polymer[29] (Figures 3d,S14, Table S5). Both $CO_2$ uptake and $CO_2/N_2$ selectivity of NPCF-10 are among the highest ones.

By fitting $CO_2$ adsorption isotherms of NPCF-10 at 273 and 298 K (Figure 3c), the isosteric heat ($Q_{st}$) is calculated to be 20–40 kJ mol$^{-1}$ depending on $CO_2$ uptake (e.g., 30.9 kJ mol$^{-1}$ at 0.5 mmol g$^{-1}$), which is higher than that of C-PIL (13–33 kJ mol$^{-1}$, Figure S15). $CO_2$ temperature programmed desorption ($CO_2$-TPD) measurement (Figure 3e) was conducted to study the intrinsic interplay between $CO_2$ molecule and carbon substrate. The PCF peak is localized between 50 and 100 °C, indicating a physisorption nature on micropores. The desorption peak maximum is moved from 68 °C (PCF) to 78 °C (NPCF-10), quantifying the effect of charge transfer to the main body of carbon fiber. The NPCF-10 peak is larger as a result of higher $CO_2$ uptake, and its desorption expands to 130 °C, characterizing the deepest binding sites. Besides, the reusability for $CO_2$ sorption is tested over 10 cycles (Figure S16). Only a slight drop of $CO_2$ uptake is observed (Figure 3f), probably due to pore blocking by high boiling-point contaminants, e.g., water. Nevertheless, its initial capacity is recovered by annealing at high temperature.

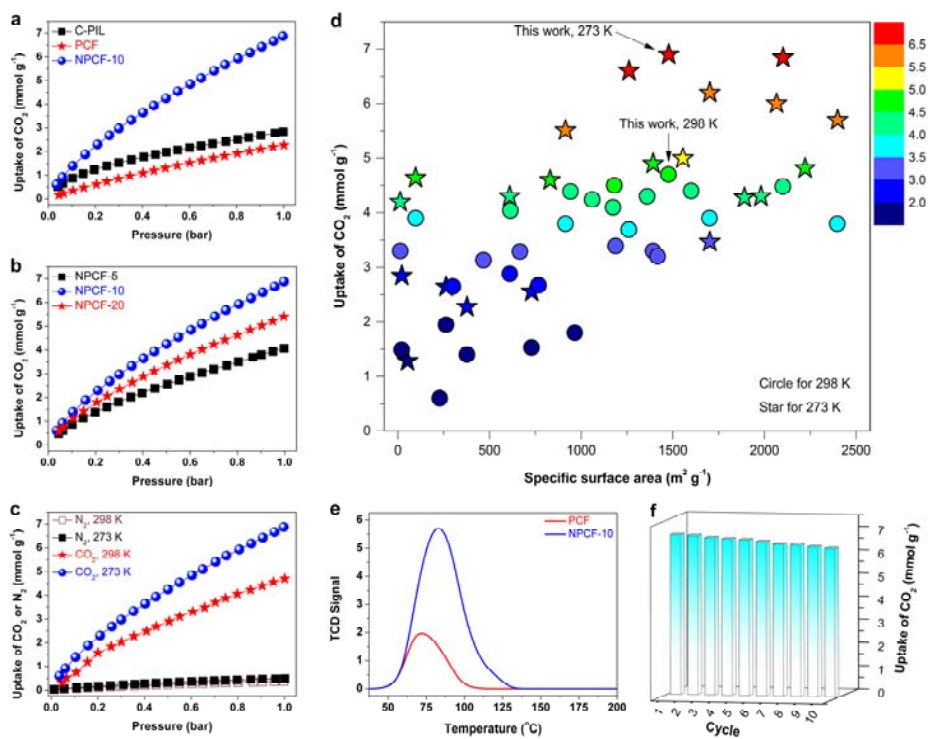

**Figure 3** (a and b) $CO_2$ adsorption isotherms at 273 K. (c) $N_2$ and $CO_2$ adsorption isotherms of NPCF-10. (d) Comparison of $CO_2$ uptake of NPCF-10 with other sorbents (Table S5). (e) $CO_2$-TPD profiles. (f) Reusability of NPCF-10 at 273 K.

The surprisingly high $CO_2$ uptake is attributed to the designed N-doped carbon/carbon biphasic heterojunction in the microporous carbon fibers. We assume that similarly to a voltaic element the more noble N-doped carbon layer takes electrons from the underlying conductive carbon and thereupon promotes its interaction with guest molecules[17]. Such carbon/carbon heterojunction is active in similar systems[16,30] and also explains their high catalytic activity[31]. The key information is presumably found in the plot of $CO_2$ adsorbed amount per specific surface area. Excluding swelling (i.e., volume uptake) as typical for polymers, this allows to calculate a "molecular footprint" or area requirement per adsorbed $CO_2$ molecule. With a value of 1476.3 m$^2$ g$^{-1}$ for NPCF-10, it is 0.36 nm$^2$, which is 40% denser than typical values for other sorbents. Assuming the bulk carbon in NPCF-10 to be positively polarized due to the heterojunction, we can only conclude that the typical quadrupolar flat $CO_2$ sorption is complemented by an upright, dipolar interaction motif working via the negative oxygen atom of $CO_2$. An experimental proof is given in Figure S17 by applying a potential to neutralize the positive charge of the bulk microporous carbon in NPCF-10 to restore the quadrupolar flat $CO_2$ sorption. A 30% drop in $CO_2$ uptake is observed, as one might expect. Summarizing all observations, the exceptionally intense $CO_2$ sorption by NPCF-10 cannot be explained by classical $CO_2$ sorption modes known for most porous carbons. A stronger and denser packing actually exists in our system.

To assess the potential of NPCF-10 in practical process, competitive $CO_2$ sorption with $N_2$ in a dynamic system was conducted using a mixed gas stream of 20% (v/v) $CO_2$ + 80% (v/v) $N_2$ (Figure S18). At 298 K and 0.2 bar partial pressure of $CO_2$, the dynamic $CO_2$ adsorption capacity is 2.1 mmol g$^{-1}$, which matches well with that from the equilibrium measurement using pure $CO_2$ (2.3 mmol g$^{-1}$). This suggests $CO_2$ preferentially adsorbs onto NPCF-10 over $N_2$ in a $CO_2/N_2$ mixture.

Resistive sensor is most attractive due to its easy fabrication and possible miniaturization[32]. To illustrate the advantage of our system beyond "only" record values of sorption, we constructed a $CO_2$ sensor based on the resistance change of the fiber upon gas exposure (Figure 4a). The $\Delta R/R_0$ is defined as the relative reduction in resistance, where $R_0$ corresponds to the original resistance of device in a multifiber array and $\Delta R$ is the reduction of resistance upon $CO_2$ exposure.

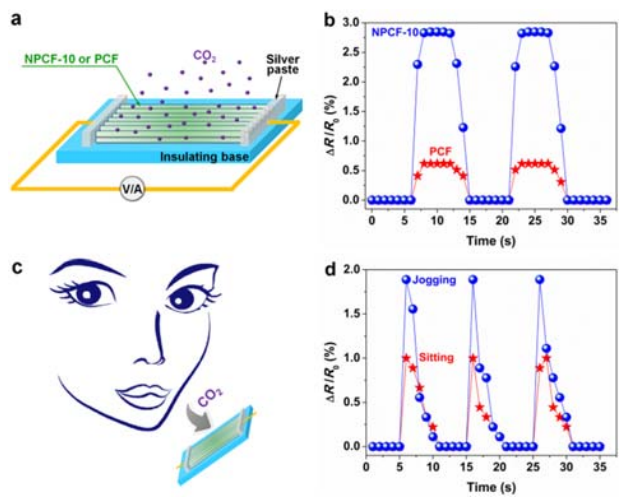

**Figure 4** (a) Scheme of $CO_2$ sensor based on NPCF-10 or PCF array. (b) Kinetic response of $CO_2$ sensor in the presence of 2.0% $CO_2$. (c) Scheme of $CO_2$ sensor for monitoring the exhale gas of human breath. (d) Kinetic response of NPCF-10-derived device to the exhale gas of human breath during sitting or jogging.

Figure 4b shows the $\Delta R/R_0$ vs. time plot of NPCF-10-based $CO_2$ sensor at a $CO_2$ concentration of 2.0%. In contact with this probe gas, the resistance decreases quickly within 2 s. We regard this as the time necessary to exchange gas throughout the complete fiber with the altered gas composition, corresponding to a diffusion coefficient of 2.45 × 10$^{-6}$ cm$^2$ s$^{-1}$. The signal of NPCF-10-based sensor ($\Delta R/R_0$ peak = 2.85%) is four times stronger than that of PCF, which reflects the improved selectivity of binding $CO_2$ over $N_2$ (please note that it is a competitive binding experiment, in which 2% $CO_2$ is sensed in the presence of 98% $N_2$, that is, a majority of surface is saturated with $N_2$). Besides, NPCF-10 shows fast recovery, outperforming many resistive sensors that require tens of seconds[33], and has little-to-no sensitivity toward humidity (Figure S19), which makes our approach superior to infrared spectrometry[11c]. As $CO_2$ concentration increases from 0.01% to 2.0%, the $\Delta R/R_0$ peak raises from 0.90% to 2.85% (Figure S20), suggesting the dynamic range is higher than that of classical carbon-based devices.

Measuring $CO_2$ level in breath allows for non-invasive, fast evaluation of perfusion and systemic metabolism, which provides a pain-free, low-cost method to diagnose asthma, chronic obstructive pulmonary disease and others at initial stage[34]. With the ability to detect $CO_2$ at ambient conditions, NPCF-10 works as humidity-tolerant $CO_2$ sensor for a simple real-time breath analysis (Figure 4c). The calm breathing of a healthy man in quiescent state yields a $\Delta R/R_0$ peak of 0.88%, while it increases to 1.89% during jogging (~10 km h$^{-1}$, Figure 4d). This primary, purely illustrative experiment result reflects the correlation of $CO_2$ concentration in the exhale gas with metabolism intensity and demonstrates the promise in condition monitoring in a time and potentially space resolved fashion.

In summary, a novel activation concept has been proposed to produce functional carbon fibers by precise deposition of nitrogen species on surface. The synthesized carbon fibers uptake $CO_2$ as much as 30% of its own mass at 273 K and 1 bar due to a contribution of $CO_2$ vertical packing in a carbon/carbon heterojunction-charge polarized system. We build up a resistive, humidity-tolerant $CO_2$ sensor operational at ambient conditions. Our synthetic protocol allows for the simultaneous optimization of surface functionalities and porous structure of carbon, and the proposed carbon/carbon heterojunction is believed to serve as a useful structural tool for solving material-based problems.

## Acknowledgements

This work was financially supported by the Max Planck Society and the European Research Council Starting Grant (639720-NAPOLI).

**Keywords:** Poly(ionic liquid) • biphasic heterojunction • carbon fiber • $CO_2$ capture • $CO_2$ sensor

# Poly(ionic liquid)-Derived Carbon with Site-Specific N-Doping and Biphasic Heterojunction for Enhanced $CO_2$ Capture and Sensing

Jiang Gong, Markus Antonietti and Jiayin Yuan*

**Abstract:** $CO_2$ capture is a pressing global environmental issue that drives scientists to develop creative strategies to tackle this challenge. The concept in this contribution is to produce site-specific nitrogen-doping in microporous carbon fibers. It creates a carbon/carbon heterojunction using poly(ionic liquid) (PIL) as "soft" activation agent that deposits nitrogen species exclusively on the skin of commercial microporous carbon fibers. Such carbon-based biphasic heterojunction amplifies the interaction between carbon fiber and $CO_2$ molecule for unusually high $CO_2$ uptake and resistive sensing.

# Table of Contents

## 1. Experimental Procedures

### 1.1 Materials and Chemicals

1-Vinylimidazole (Sigma-Aldrich, purity ≥99%), bromoacetonitrile (Sigma-Aldrich, purity ≥ 97%), 2,2'-azobis(2-methylpropionitrile) (AIBN, Sigma-Aldrich, purity ≥ 98%), and bis(trifluoromethane sulfonyl)imide lithium salt (LiTf$_2$N, Sigma-Aldrich, purity ≥ 99%) were used as received without further purifications. Diethyl ether, dimethyl sulfoxide (DMSO), *N*,*N*-dimethylformamide (DMF), acetate and methanol were of analytic grade. Commercial microporous carbon fibers (PCFs) with the average diameter of ca. 25.4 μm and length of 4−6 cm were kindly provided by Anshan Sinocarb Carbon Fibers Co., Ltd.

### 1.2 Synthesis of Poly(ionic liquid) (PIL)

Poly(3-cyanomethyl-1-vinylimidazolium bromide) (PCMVImBr) was firstly synthesized (Figure S1). Briefly, 20.0 g of the monomer 3-cyanomethyl-1-vinylimidazolium bromide (CMVImBr, prepared from the reaction of 1-vinylimidazole and bromoacetonitrile in diethyl ether at room temperature for 48 h), 400 mg of AIBN, and 200 mL of DMSO were loaded into a 500 mL of reactor. The mixture was deoxygenated several times by a freeze-pump-thaw procedure. The reactor was then refilled with nitrogen and placed in an oil bath at 70 °C for 24 h. The obtained mixture was exhaustively dialyzed against water (the molecular weight cut-off of the dialysis bag is 8 kDa) for one week and then freeze-dried from water. The gel permeation chromatography (GPC) trace of the product PCMVImBr is shown in Figure S2.

Subsequently, poly[3-cyanomethyl-1-vinylimidazolium bis(trifluoromethane sulfonyl)imide] (PCMVImTf$_2$N, simplified as "PIL") was synthesized *via* anion exchange with PCMVImBr using LiTf$_2$N salt in aqueous solution. The $^1$H nuclear magnetic resonance ($^1$H NMR) spectrum of the as-prepared PIL is displayed in Figure S3.

### 1.3 Synthesis of NPCF

Firstly, a defined amount of PIL was dissolved in 100 mL of DMF to form a homogeneous solution. Subsequently, a defined amount of PCF was completely immersed in the PIL solution under gentle agitation for 12 h. The PCF@PIL-*x* composite (*x* = 5, 10 and 20, *x* denotes the mass ratio of PIL to PCF in percentage when mixed together in DMF; it should be noted that a part of PIL chains were attached onto the surface of PCF and the other PILs still remained in DMF solution) was obtained by separation, solvent evaporation at 80 °C, and drying in a vacuum oven at 100 °C for 24 h to evaporate the residual DMF.

The PIL content of PCF@PIL-*x* composite (*x* = 5, 10 and 20) was measured by weighing method to be ca. 0.6, 0.8 and 1.1 wt %, respectively. We also checked the real content of PIL by measuring the nitrogen content of PCF@PIL-*x* composite (since the nitrogen content of PCF is zero, and all nitrogen atoms in the composite come from the PIL) through combustion elemental analyses. In this way, the PIL content of the PCF@PIL-*x* composite was calculated to be ca. 0.8, 0.9 and 1.3 wt %, respectively, which is close to the results obtained by weighing the fiber before and after the coating step.

Afterwards, the resultant PCF@PIL-*x* composite was firstly heated to 550 °C for 1 h, and subsequently calcined at 750 °C for 1 h under N$_2$ atmosphere using a Nabertherm N7/H chamber oven with a P300 controller. The heating rate was kept at 10 °C min$^{-1}$. After slowly cooling down to the room temperature, NPCFs were obtained and denoted as NPCF-*x* (*x* = 5, 10 and 20, *x* is defined in the PCF@PIL-*x* composite). In comparison, the powder carbon product prepared from the carbonization of PIL in the absence of PCF substrate under similar fabrication process was denoted as C-PIL. Besides, to study the carbonization conditions on the physical and chemical nature of NPCF-10, other carbonization temperatures (650 and 850 °C), carbonization times (0.5 and 2 h) and heating rates (5 and 20 °C min$^{-1}$) are also applied.

### 1.4 CO$_2$ Capture and Sensor

CO$_2$ capture by C-PIL, PCF and NPCFs was accomplished by measuring the adsorption isotherms at 273 or 298 K in a Quantachrome Autosorb and Quadrasorb (Quantachrome Instruments). Before each analysis, the samples were degassed at 150 °C for 24 h. The isosteric heat ($Q_{st}$) of CO$_2$ adsorption was calculated by applying the Clausius−Clapeyron equation to the adsorption isotherms measured at the aforementioned two temperatures:

$$\ln(\frac{p_1}{p_2}) = Q_{st} \times \frac{T_2 - T_1}{R \times T_1 \times T_2} \tag{1}$$

where $p_1$ and $p_2$ denote the pressure (Pa), $R$ is the universal gas constant (8.314 J mol$^{-1}$ K$^{-1}$), and $T_1$ and $T_2$ represent the absolute temperature (K).

To investigate the effect of charge on the CO$_2$ uptake by NPCF-10, 2.0 g of NPCF-10 was placed on a plastic carrier of a lab electronic balance. The whole balance was sealed, and gas inlet and outlet were kept open. Briefly, N$_2$ gas was firstly purged into the device with the flow velocity of 100 mL min$^{-1}$ for 1 h. The reading of the balance became stable and was then cleared to zero. Afterwards, the CO$_2$ gas (100 mL min$^{-1}$) was then purged into the device. The reading increased fast at first due to the adsorption of CO$_2$ by NPCF-10, and then slowly reached a maximum value. When NPCF-10 was stimulated by a battery (9 V), the reading number was observed to decrease moderately.

The breakthrough experiment of NPCF-10 was performed in a small-scale fixed-bed column. The feed stream with a composition of CO$_2$/N$_2$ (20% of CO$_2$ in volume) was fed into the column at a flow rate of 20 mL min$^{-1}$. Prior to the sorption experiment, the sample was heated to 200 °C under N$_2$ flow at 100 mL min$^{-1}$ for 4 h to desorb adventitious CO$_2$ and water, slowly cooled to 25 °C and then exposed to CO$_2$ for the experimental sorption run.

To explore the performance of NPCF-10 (or PCF)-derived device (shown in Figure 4a) as the CO$_2$ gas sensor, we measured the resistance change of the device upon exposure to CO$_2$ gas at 25 °C. The fabricated NPCF-10 (or PCF)-derived device was placed inside a chamber connected to a two-input mixer that mixes CO$_2$ gas and N$_2$ gas. The resistance of the device at a CO$_2$ concentration ranging from 0.01% to 2.0% was measured by using Digital-Multimeter Benning MM 7-1. The response time is defined as the time taken for the relative resistance change to reach 90% of the steady-state value. The recovery time is defined as the time needed to recover to 10% of the original resistance.

To study the effect of humidity on the resistance change of the device, the NPCF-10-based CO$_2$ sensor was placed in an environmental chamber (model WKL 34, Weiss Technik), whereby the temperature was kept at ca. 25 °C and the relative humidity was modulated from 20% to 90%, during which the resistance of the NPCF-10-based CO$_2$ sensor was recorded.

**1.5 Characterization**

GPC measurement was conducted at 25 °C on NOVEMA-column with mixture of 80% acetate buffer and 20% methanol as eluent (flow rate = 1.0 mL min$^{-1}$, PEO standards using RI detector-Optilab-DSP-Interferometric Refractometer). $^1$H NMR measurement using DMSO-$d_6$ as the solvent was carried out at the room temperature using a Bruker DPX-400 spectrometer operating at 400 MHz. Scanning electron microscopy (SEM) measurements were carried out in a LEO 1550-Gemini electron microscope (acceleration voltage = 3 kV), and the samples were coated with a thin gold layer before SEM measurements. Energy-dispersive X-ray (EDX) maps were taken on the SEM with an EDX spectrometer. Combustion elemental analyses were done with a varioMicro elemental analysis instrument from Elementar Analysensysteme. The surface element composition was characterized by means of X-ray photoelectron spectroscopy (XPS) carried out on a VG ESCALAB MK II spectrometer using an Al K$_\alpha$ exciting radiation from an X-ray source operated at 10.0 kV and 10 mA.

N$_2$ adsorption/desorption experiments were performed with a Quantachrome Autosorb and Quadrasorb at 77 K, and the data were analyzed using Quantachrome software. The specific surface area was calculated by using Brunauer-Emmett-Teller (BET) equation. The pore voulme was calculated by using *t*-method. The pore size distribution was obtained by applying the quench solid density functional theory (QSDFT) on the adsorption branch and assuming slit-like geometry on carbon material kernel. The samples were degassed at 150 °C for 24 h before measurements. X-ray diffraction (XRD) patterns were recorded on a Bruker D8 diffractometer using Cu K$_\alpha$ radiation ($\lambda$ = 0.154 nm) and a scintillation counter. Raman spectra were collected using a confocal Raman microscope ($\alpha$300; WITec, Ulm, Germany) equipped with a 532 nm laser. The electrical conductivity was measured using Benning MM 7-1multimeter. Mechanical properties were measured on an Instron 1121 at an extension speed of 1 mm min$^{-1}$. All data were the average of five independent measurements; the relative errors committed on each data were reported as well.

CO$_2$ temperature programmed desorption (CO$_2$-TPD) measurements were carried out on a Micromeritics Chemisorption Analyzer (USA). Before the measurement, the sample was firstly pre-treated at 250 °C for 2 h under helium atmosphere. After the temperature slowly decreased to 25 °C, the sample was swept by $CO_2$ for 2 h, and then the gas was switched to helium until the baseline was flat. Subsequently, the temperature was gradually increased to 250 °C with a ramping rate of 10 °C min$^{-1}$ to obtain the $CO_2$-TPD curve. Generally, the peak below ca. 100 °C in the $CO_2$-TPD profile is mainly owing to physical adsorption of $CO_2$ molecule by porous structure. The peak for the desorption temperature above ca. 100 °C is usually believed to be related to chemical adsorption of $CO_2$ molecule.

## 2. Supporting Figures and Tables

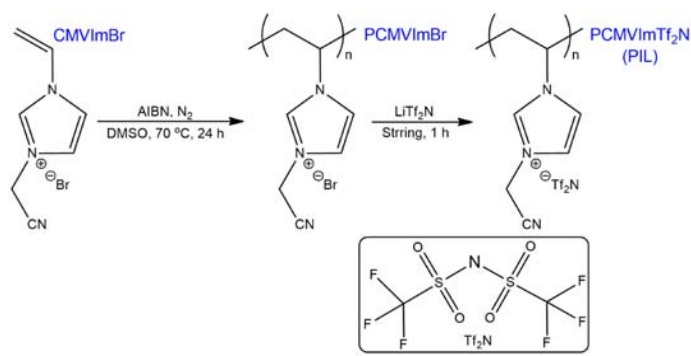

**Figure S1** Synthetic route to the PCMVImTf$_2$N (simplified as "PIL" in the manuscript).

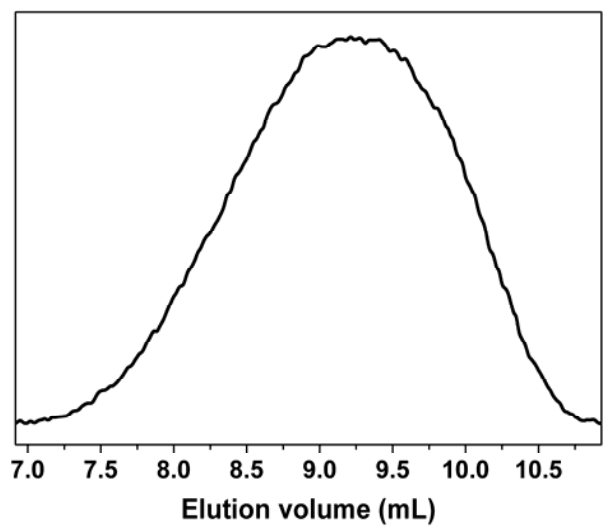

**Figure S2** GPC trace of the PCMVImBr.

The apparent number-average molecular weight and polydispersity index value of the PCMVImBr are measured to be $1.97 \times 10^5$ g mol$^{-1}$ and 2.67, respectively. Since the PIL is synthesized by anion exchange of the PCMVImBr with LiTf$_2$N salt in aqueous solution, the apparent number-average molecular weight of the PIL is calculated to be $3.80 \times 10^5$ g mol$^{-1}$.

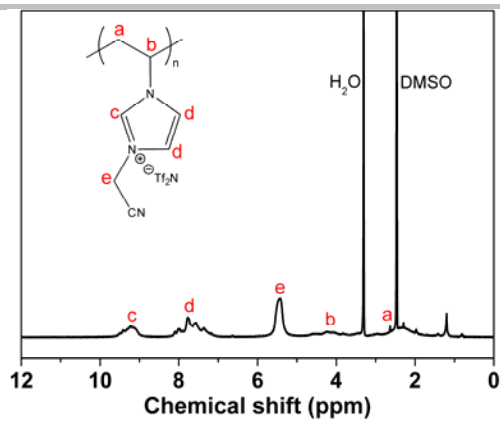

**Figure S3** $^1$H NMR spectrum of the PIL using DMSO-$d_6$ as the solvent.

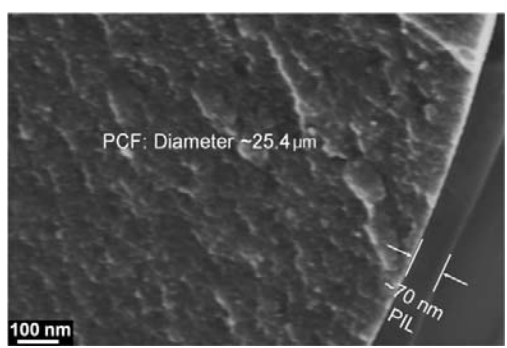

**Figure S4** A representative SEM image of the cross-sectional structure of the core/sheath PCF@PIL-20 composite before carbonization. The thickness of the supported PIL layer in the PCF@PIL-x composite ranges from 30 to 70 nm depending on the initial mass ratio of PIL to PCF (*x*) when mixed in DMF.

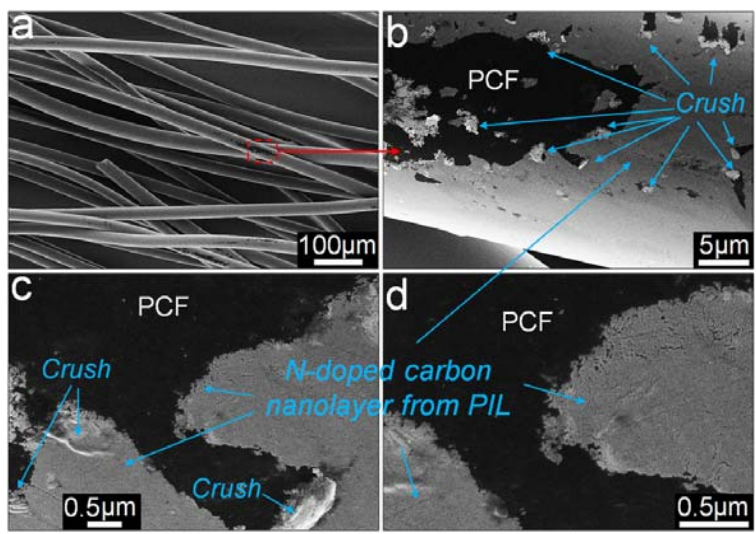

**Figure S5** (a–d) Typical SEM images of NPCF-10 surface structure after being gently scraped by tweezers to clearly visualize the surface nitrogen-doped carbon nanolayer. These crushes in (b–c) indicate the presence of the porous carbon nanolayer coating on the surface of NPCF-10.

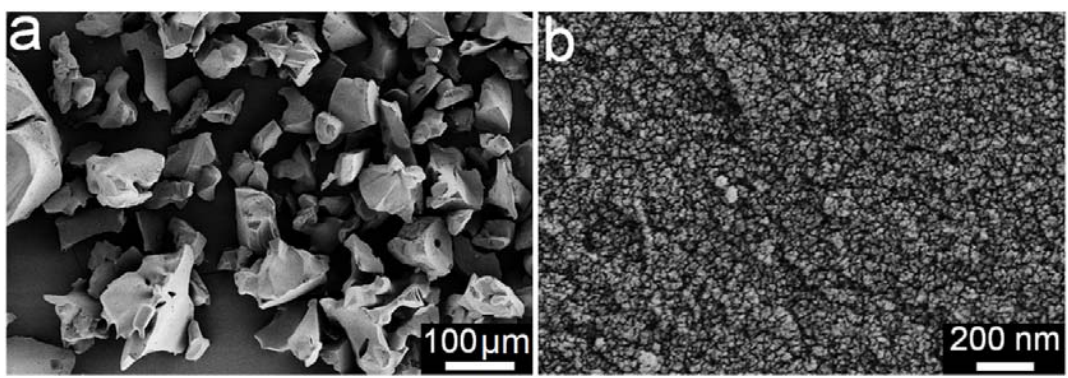

**Figure S6** Typical SEM images with (a) low and (b) high magnifications of the C-PIL prepared from the carbonization of merely PIL in the absence of PCF substrate at 750 °C. Evidently, C-PIL contains many large macroparticles with the size of ca. 50–200 μm. Each large macroparticle consists of numerous aggregated nanoparticles with the size of ca. 30–50 nm.

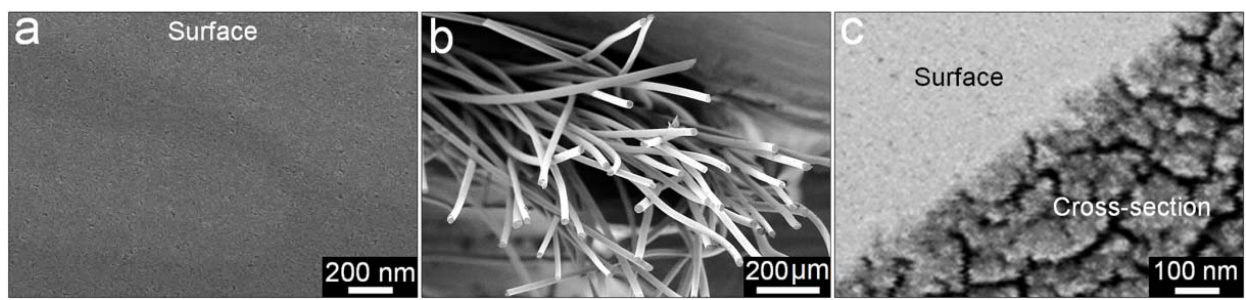

**Figure S7** Typical SEM images of (a) surface structure, and (b and c) cross-sectional structure of raw PCF. As we can see, the surface of raw PCF is smooth and closed.

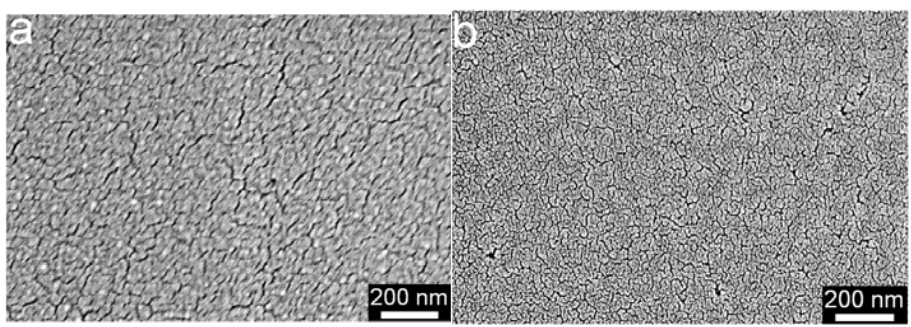

**Figure S8** Typical SEM images of the surface microstructures of (a) the native PIL, and (b) the supported PIL thin layer on the carbon fiber surface in the PCF@PIL-10 composite before carbonization.

The sizes of the strip-shaped patches in the native PIL and the supported PIL thin layer on the surface of PCF are ca. 40–80 and 25–40 nm, respectively. The widths of their nanocanyons between these patches are ca. 10–15 and 5–10 nm, respectively.

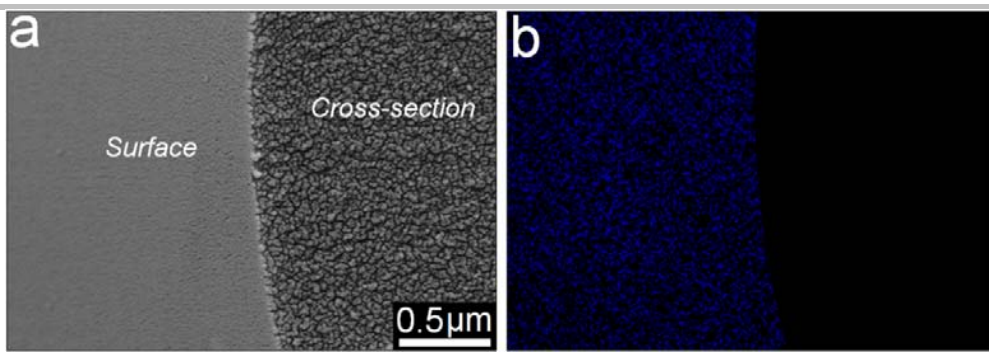

**Figure S9** (a) Typical SEM image and (b) nitrogen EDX map of the cross-sectional structure of NPCF-10. The nitrogen EDX mapping of NPCF-10 verifies the uniform distribution of nitrogen (blue area on the left in (b)) in the porous surface carbon nanolayer rather than the inside of the porous carbon fiber.

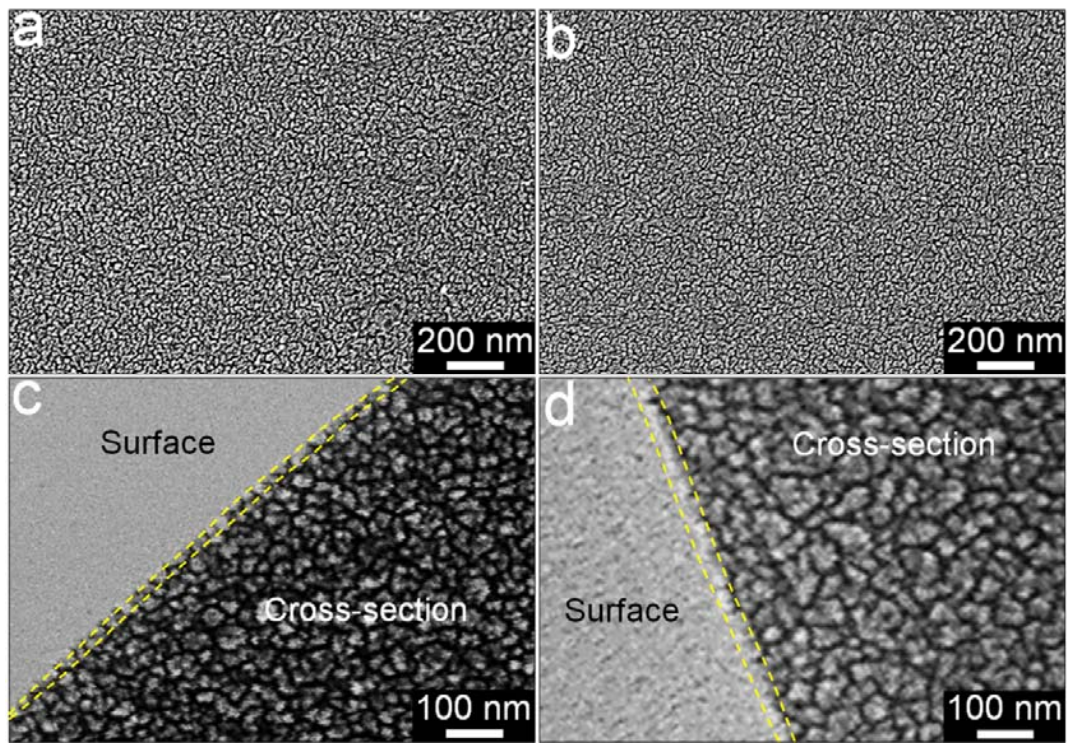

**Figure S10** Typical SEM images of NPCF-5 (a for surface structure and c for cross-sectional structure) and NPCF-20 (b for surface structure and d for cross-sectional structure). Similar to NPCF-10, both NPCF-5 and NPCF-20 also display the core/sheath structure, consisting of a porous carbon fiber core and a PIL-derived porous nitrogen-doped carbon layer sheath.

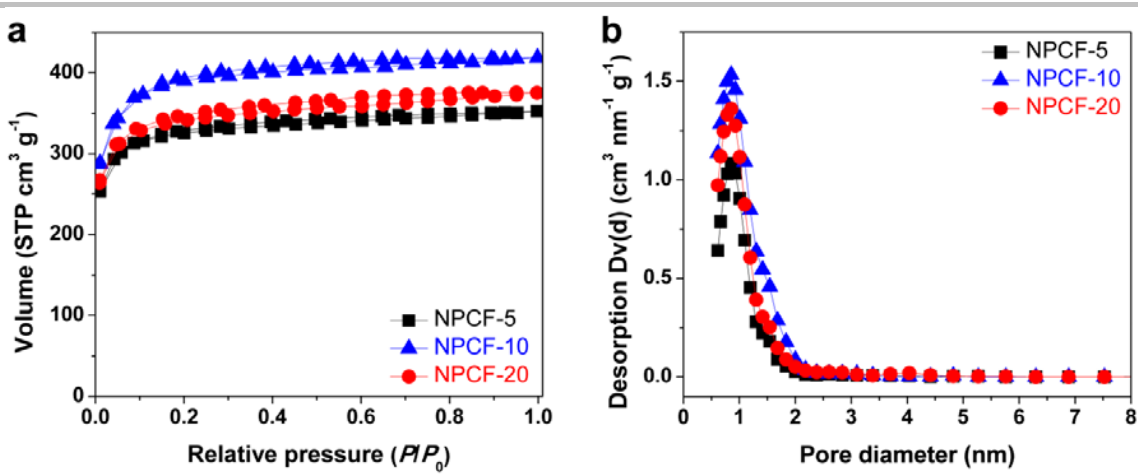

**Figure S11** (a) $N_2$ adsorption/desorption isotherms at 77 K and (b) pore size distribution plots of NPCFs.

As we can see, all of NPCFs deliver type I physisorption isotherm, and their corresponding pore size distributions are ranging from 0.6 to 2.2 nm.

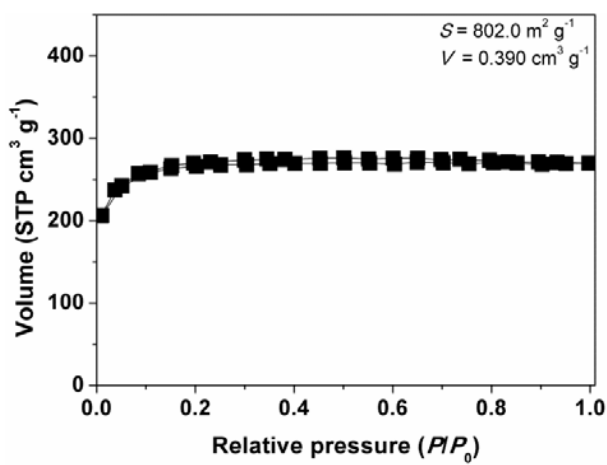

**Figure S12** $N_2$ adsorption/desorption isotherms at 77 K of the physical mixture of C-PIL (0.4 wt %) and PCF (99.6 wt %).

Obviously, the $S$ (802.0 m² g⁻¹) and $V$ (0.390 cm³ g⁻¹) of the physical mixture of C-PIL and PCF are lower than those of NPCF-10 ($S$ = 1476.3 m² g⁻¹, and $V$ = 0.583 cm³ g⁻¹).

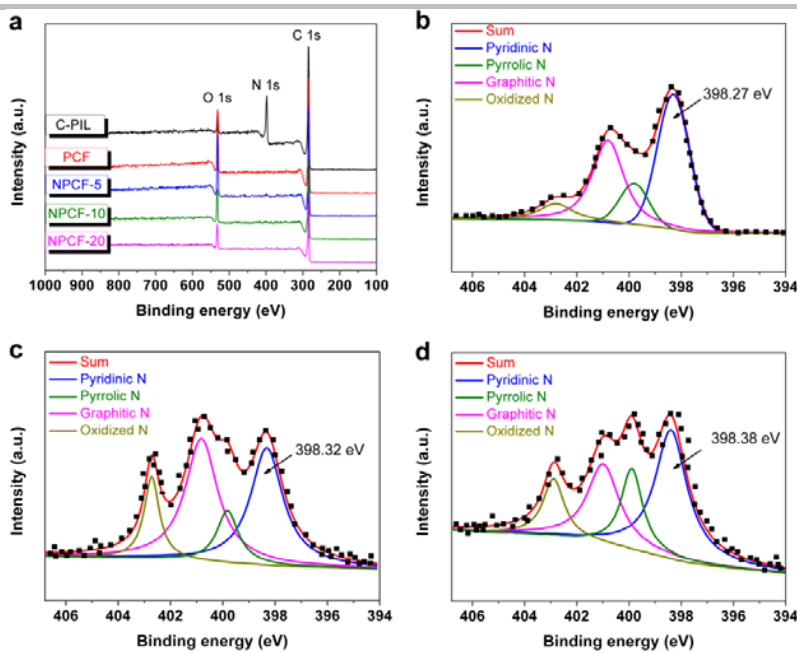

**Figure S13** (a) XPS spectra of C-PIL, PCF and NPCFs, with C 1s (ca. 284.6 eV), N 1s (ca. 398.4 eV) and O 1s (ca. 532.3 eV) peaks. High-resolution N 1s XPS spectra of (b) C-PIL, (c) NPCF-5, and (d) NPCF-20.

High-resolution N 1s XPS peak is deconvoluted into four components: pyridinic N (ca. 398.3 eV), pyrrolic N (ca. 399.8 eV), graphitic N (ca. 400.9 eV), and oxidized N (ca. 402.8 eV), according to the previous work[1]. The detailed results are listed in Table S2. The shift of the pyridinic N peak of NPCF-10 (at 398.37 eV, see Figure 2e) with regard to that of C-PIL (at 398.27 eV, see Figure S13b) is ca. 0.10 eV, comparable to that in a previous report (ca. 0.09 eV)[2]. This shift is supposed to be owing to the charge transfer in the carbon/carbon heterojunction of NPCF-10[2,3].

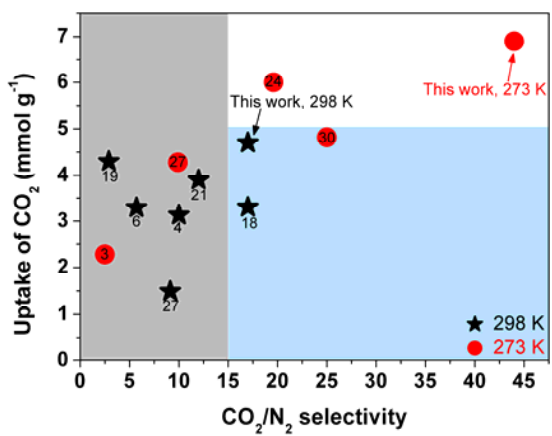

**Figure S14** Comparison of the apparent $CO_2/N_2$ selectivity of NPCF-10 with other sorbents.

The number near to the circle or star indicates the entry number listed in Table S5. As we can see, NPCF-10 shows relatively high apparent $CO_2/N_2$ selectivities at 273 and 298 K.

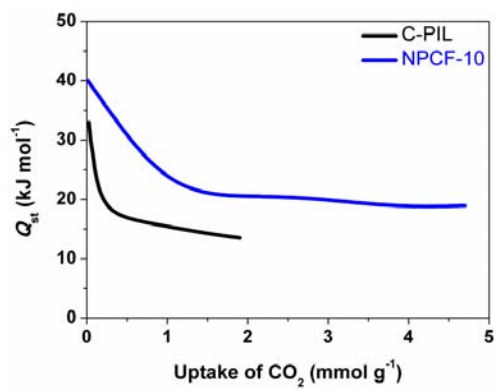

**Figure S15** Isosteric heat ($Q_{st}$) of $CO_2$ adsorption by C-PIL and NPCF-10 at different $CO_2$ uptake amounts.

The $Q_{st}$ value of NPCF-10 is calculated to be in the range of 20–40 kJ mol$^{-1}$ (*e.g.*, 30.9 kJ mol$^{-1}$ at 0.5 mmol g$^{-1}$, and 20.6 kJ mol$^{-1}$ at 2.0 mmol g$^{-1}$). Therefore, the $Q_{st}$ value of NPCF-10 is higher than that of C-PIL (13–33 kJ mol$^{-1}$), which suggests the effect of the heterojunction in the NPCF-10.

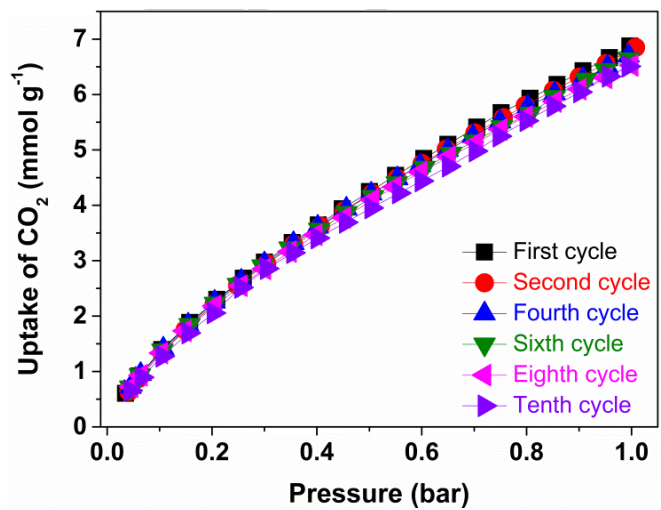

**Figure S16** $CO_2$-multicircle-sorption isotherms for the NPCF-10 at 273 K over 10 cycles.

After 10 cycles, the $CO_2$ adsorption isotherm of NPCF-10 at 273 K changes slightly, but the $CO_2$ uptake is still up to 6.4 mmol g$^{-1}$, indicating its good reusability for $CO_2$ capture. Importantly, after 10 cycles, this initial capacity is regained *via* regeneration treatment (*e.g.*, 180 °C for 24 h under high vacuum).

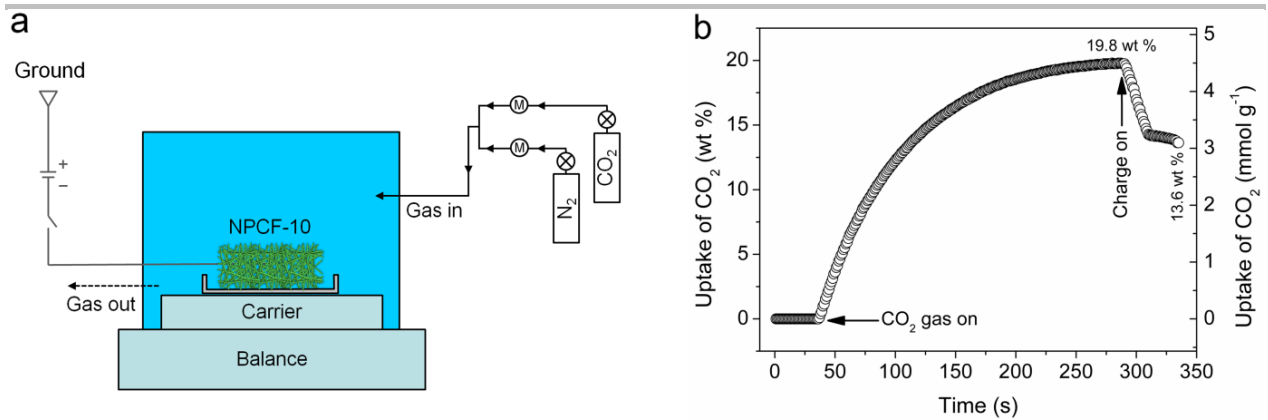

**Figure S17** Model experiment analyzing the charge function in $CO_2$ uptake. (a) Scheme of the model device to explore the effect of charge on $CO_2$ uptake by NPCF-10. (b) The plot of $CO_2$ uptake by NPCF-10 vs. time at 298 K.

When an electric current depends on $CO_2$-sorption, $CO_2$-sorption usually will depend on electric current. This in principle enables the electric recovery of binding sites, that is, one can envision cyclic processes where a carbon filter fabric binds $CO_2$, while in a second cyclic stroke the $CO_2$ is released by an electric current. This option is extremely simpler than the currently applied pressure or external temperature swing process. We applied this principle to turn an ordinary lab electronic balance into a $CO_2$ separation machine. The "device" first takes up 19.8 wt % of NPCF-10 weight of $CO_2$ at 298 K, while releasing about a third of it after being stimulated by an added potential of 9 V. The released amount is what we are able to put in extra by the heterojunction effect. This effect vanishes after about 10 cycles, which we attribute to the onset of a number of electrochemical reactions such as oxidation of surface nitrogen species.

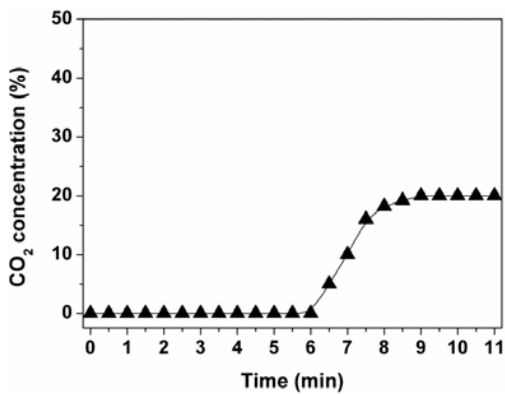

**Figure S18** Breakthrough curve for NPCF-10 obtained at 298 K and 1 bar.

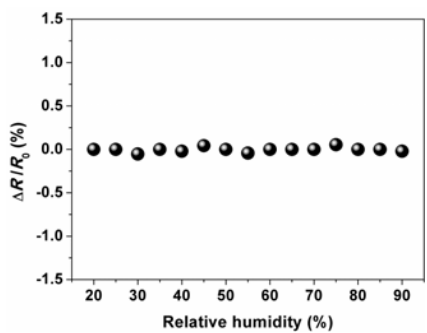

**Figure S19** The resistance change of the NPCF-10-derived $CO_2$ sensor at the relative humidity of 20%–90%.

The NPCF-10-derived $CO_2$ sensor shows a little-to-no sensitivity resistance change when the relative humidity increases from 20% to 90% at 25 °C, indicating that NPCF-10 is immune to water vapor. That is to say, no filtering is needed for the pre-purification of $CO_2$ prior to monitoring by NPCF-10, suggesting the advantage of our approach over the traditional infrared detection technology.

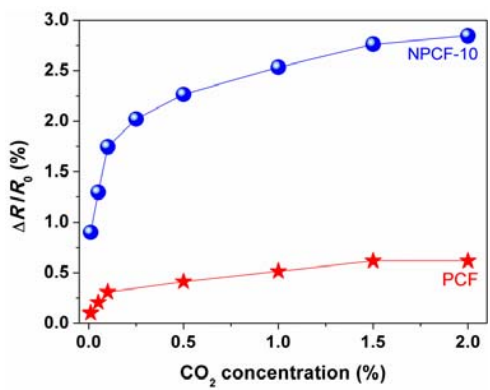

**Figure S20** The resistance change of the NPCF-10 (or PCF)-derived $CO_2$ sensor at different $CO_2$ concentrations.

**Table S1** Conversion of PIL, and element composition of C-PIL, PCF, and NPCFs.

| Entry | Sample | Conversion of PIL in carbonization (wt %) [a] | C [b] (wt %) | N [b] (wt %) | H [b] (wt %) | O [c] (wt %) |
|---|---|---|---|---|---|---|
| 1 | C-PIL | 21.7 | 64.7 | 14.51 | 2.2 | 18.6 |
| 2 | PCF | - | 84.3 | 0 | 1.4 | 14.3 |
| 3 | NPCF-5 | 39.2 | 82.5 | 0.21 | 1.2 | 16.1 |
| 4 | NPCF-10 | 45.1 | 85.7 | 0.32 | 1.4 | 12.6 |
| 5 | NPCF-20 | 47.0 | 85.1 | 0.42 | 1.6 | 12.9 |

[a] Calculated by the mass ratio of the PIL-derived carbon fraction to the PIL in the composite before carbonization. [b] Measured by combustion element analyses. [c] Calculated by the difference.

The improved conversion of PIL on the surface of PCF is consistent with the result recently reported by Su's group[4]. They prove that the strong interaction between ionic liquids (*i.e.*, the monomer of PILs) and graphitic support allows the entrapping and directing of ionic liquids on the surface of graphitic support, which thus facilitates the carbonization of ionic liquids.

**Table S2** Textural parameter and $CO_2$ uptake of C-PIL, PCF, and NPCFs.

| Entry | Sample | $S$ [a] (m² g⁻¹) | $S_{micro}$ [b] (m² g⁻¹) | $V$ [c] (cm³ g⁻¹) | $V_{micro}$ [d] (cm³ g⁻¹) | $D_A$ [e] (nm) | $CO_2$ uptake (mmol g⁻¹) |
|---|---|---|---|---|---|---|---|
| 1 | C-PIL | 707.6 | 523.8 | 0.382 | 0.275 | 1.2 | 2.8 [f] |
| 2 | PCF | 800.8 | 750.4 | 0.396 | 0.378 | 0.8 | 2.3 [f] |
| 3 | NPCF-5 | 1024.2 | 965.6 | 0.474 | 0.453 | 0.8 | 4.1 [f] |
| 4 | NPCF-10 | 1476.3 | 1402.0 | 0.583 | 0.560 | 0.8 | 6.9 [f]/4.7 [g] |
| 5 | NPCF-20 | 1319.0 | 1251.0 | 0.523 | 0.497 | 0.8 | 5.4 [f] |

[a] BET specific surface area. [b] Specific surface area of micropores. [c] Pore volume. [d] Volume of micropores. [e] Average diameter of pores. [f] $CO_2$ uptake at 273 K. [g] $CO_2$ uptake at 298 K.

**Table S3** Surface element composition of C-PIL, PCF, and NPCFs measured by XPS.

| Entry | Sample | C (at %) | O (at %) | N (at %) | Pyridinic N (%) | Pyrrolic N (%) | Graphitic N (%) | Oxidized N (%) |
|---|---|---|---|---|---|---|---|---|
| 1 | C-PIL | 74.9 | 17.3 | 7.8 | 43.9 | 12.4 | 26.4 | 7.3 |
| 2 | PCF | 82.5 | 17.5 | 0 | 0 | 0 | 0 | 0 |
| 3 | NPCF-5 | 79.7 | 18.7 | 1.6 | 37.4 | 11.2 | 39.2 | 12.2 |
| 4 | NPCF-10 | 84.6 | 13.6 | 1.8 | 31.6 | 22.9 | 29.7 | 15.8 |
| 5 | NPCF-20 | 83.1 | 12.9 | 3.0 | 42.7 | 18.2 | 26.4 | 12.7 |

**Table S4** Effect of carbonization conditions (*i.e.*, carbonization temperature, time and heating rate) on the physical and chemical nature of NPCF-10 obtained from the carbonization of PCF@PIL-10 composite.

| Entry | Temperature (°C) | Time (h) | Heating rate (°C min$^{-1}$) | $S$ [a] (m$^2$ g$^{-1}$) | $I_G/I_D$ [b] | C [c] (wt %) | N [c] (wt %) | H [c] (wt %) | O [d] (wt %) |
|---|---|---|---|---|---|---|---|---|---|
| 1 | 650 | 1 | 10 | 1402.4 | 0.92 | 86.2 | 0.39 | 1.7 | 11.7 |
| 2 | 750 | 1 | 10 | 1476.3 | 0.91 | 85.7 | 0.32 | 1.4 | 12.6 |
| 3 | 850 | 1 | 10 | 1509.5 | 0.89 | 84.9 | 0.21 | 1.1 | 13.8 |
| 4 | 750 | 0.5 | 10 | 1400.7 | 0.91 | 86.0 | 0.38 | 1.6 | 12.0 |
| 5 | 750 | 2 | 10 | 1489.2 | 0.90 | 85.2 | 0.22 | 1.2 | 13.4 |
| 6 | 750 | 1 | 5 | 1480.5 | 0.92 | 86.0 | 0.31 | 1.4 | 12.3 |
| 7 | 750 | 1 | 20 | 1468.6 | 0.91 | 85.8 | 0.33 | 1.4 | 12.5 |

[a] Specific surface area calculated from the BET equation. [b] Intensity ratio of G band to D band in Raman spectrum. [c] Measured by combustion element analyses. [d] Calculated by the residue mass.

As we can see from Entry 1–3, when the carbonization temperature raises from 650 to 850 °C, the nitrogen content of NPCF-10 decreases from 0.39 to 0.21 wt %, meanwhile its specific surface area increases from 1402.4 to 1509.5 m$^2$ g$^{-1}$. This is possibly because the increasing temperature promotes the activation process, and high temperature may accelerate the decomposition process of nitrogen species.

Similarly, when the carbonization time increases from 0.5 to 2 h (Entry 2, 4 and 5), the nitrogen content decreases from 0.38 to 0.22 wt %, and at the same time the specific surface area increases from 1400.7 to 1489.2 m$^2$ g$^{-1}$. This can be attributed to the enhanced activation effect by extended carbonization time; however the long carbonization time lowers down the nitrogen content.

Besides, as shown in Entry 2, 6 and 7, the heating rate does not obviously influence the nitrogen content and specific surface area of NPCF-10 under our experimental conditions. Since the (002) diffraction peak in the XRD pattern of NPCF-10 is weak and broad (Figure 2g), we use Raman spectroscopy to study the crystallite structure. The $I_G/I_D$ value of NPCF-10 varies slightly under different conditions, suggesting little-to-no change of the crystallite structure. Based on the results above, it can be concluded that the carbonization conditions do influence the physical and chemical nature of NPCF-10 but only to a certain extent.

Table S5 Comparisons of specific surface area ($S$), apparent $CO_2/N_2$ selectivity, and $CO_2$ uptake of NPCF-10 with a series of adsorbents reported previously.

| Entry | Post-treatments or experiment conditions | Adsorbent | $S$ (m$^2$ g$^{-1}$) | $CO_2/N_2$ (298 K) | $CO_2$ uptake (mmol g$^{-1}$, 298 K) | $CO_2/N_2$ (273 K) | $CO_2$ uptake (mmol g$^{-1}$, 273 K) | Ref. in SI |
|---|---|---|---|---|---|---|---|---|
| 1 | No post-treatments | N-doped carbon monolith | 13 | -[a] | 3.3 | - | 4.2 | [5] |
| 2 | No post-treatments | N-doped microporous carbon | 263 | - | 1.95 | - | 2.65 | [6] |
| 3 | No post-treatments | Microporous carbonaceous material | 377 | - | 1.4 | 2.5 | 2.28 | [7] |
| 4 | No post-treatments | N-doped porous carbon monolith | 467 | 10 | 3.13 | - | - | [8] |
| 5 | No post-treatments | Porous carbon nanosheet | 610 | - | 2.88 | - | 4.3 | [9] |
| 6 | No post-treatments | N-doped hierarchical carbon | 666 | 5.7 | 3.29 | - | - | [10] |
| 7 | No post-treatments | Microporous carbon material | 1174 | - | 4.1 | - | - | [11] |
| 8 | No post-treatments | Hierarchically porous carbon | 829 | - | - | - | 4.6 | [12] |
| 9 | No post-treatments | Nanoporous silicon carbide-derived carbon | 1554 | - | - | - | 5 | [13] |
| 10 | SiO$_2$ sphere as template | N-doped hollow carbon nanosphere | 767 | - | 2.67 | - | - | [14] |
| 11 | Silica SBA-15 as template KOH activation/600 °C/1 h | N-doped microporous carbon | 614 | - | 4.04 | - | - | [15] |
| 12 | KOH activation/700 °C/1 h | N-doped porous carbon material | 914 | - | 3.8 | - | 5.51 | [16] |
| 13 | KIT-6 silica as a template | Porous N-doped carbonaceous adsorbent | 942 | - | 4.39 | - | - | [17] |
| 14 | KOH activation/700 °C/0.5 h | N-doped activated carbon | 1060 | - | 4.24 | - | - | [18] |
| 15 | KOH activation/600 °C/2 h | Porous carbon | 1260 | - | - | - | 6.6 | [19] |
| 16 | Pluronic P-123 as template | Polybenzoxazine-based monodisperse carbon sphere | 1188 | - | 3.39 | - | 5.16 | [20] |
| 17 | KOH activation/600 °C/1 h | N-doped mesoporous carbon | 1260 | - | 3.69 | - | - | [21] |
| 18 | $CO_2$ activation/900 °C/3 h | Porous carbon monolith | 1392 | 17 | 3.3 | - | 4.9 | [22] |
| 19 | KOH activation/600 °C/1 h | N-doped porous carbon | 1360 | 2.9 | 4.3 | - | - | [23] |
| 20 | F127 as template KOH activation/600 °C/1 h | N-doped ordered mesoporous carbon | 1417 | - | 3.2 | - | - | [24] |
| 21 | KOH activation/600 °C/2 h | N-doped porous carbon | 1700 | 12 | 3.9 | - | 6.2 | [25] |
| 22 | Pluronic P-123 as template | N-doped porous carbon | 1979 | - | - | - | 4.3 | [26] |
| 23 | KOH activation/700 °C/2 h | Porous Na-impregnated N-doped carbon | 2100 | - | 4.48 | - | 6.84 | [27] |
| 24 | KOH activation/700 °C/2 h | Activated carbon material | 2065 | - | - | 19.6 | 6 | [28] |
| 25 | Zeolite-type MOF as template | Nitrogen decorated nanoporous carbon | 2397 | - | 3.8 | - | 5.7 | [29] |
| 26 | Polyhedral oligomeric | Well-defined | 1889 | - | - | 9.9 | 4.28 | [30] |

| | | | | | | | |
|---|---|---|---|---|---|---|---|
| | silsesquioxane as template | microporous carbon | | | | | |
| 27 | - | Supramolecular organic framework | 21 | 9.1 | 1.49 | - | 2.85 | [31] |
| 28 | - | Crystalline mesoporous germanate | 228 | - | 0.6 | - | - | [32] |
| 29 | - | Conjugated microporous polymer aerogel | 1701 | - | - | - | 3.47 | [33] |
| 30 | - | Microporous polycarbazole | 2220 | - | - | 25 | 4.82 | [34] |
| 31 | - | Metal-organic framework (SIFSIX-3-Cu) | 300 | - | 2.65 | - | - | [35] |
| 32 | - | Metal-organic framework nanosheet | 53 | - | - | - | 1.29 | [36] |
| 33 | - | Supramolecular organic framework | 97 | - | 3.9 | - | 4.64 | [37] |
| 34 | - | Nanoporous covalent organic polymer | 729 | - | 1.53 | - | 2.56 | [38] |
| 35 | - | Cobalt-coordinated conjugated microporous polymer | 965 | - | 1.8 | - | - | [39] |
| 36 | - | Metal-organic framework | - | - | - | - | 4.46 | [40] |
| 37 | PIL as "soft" activation agent/no post-treatments [b] | NPCF-10 | 1476.3 | 17 | 4.7 | 44 | 6.9 | This work |

[a] Not given/available. [b] PIL is used as a new mild porosion/activation agent, and after the carbonization of the core/sheath PCF@PIL composite, no post-treatments or purifications are needed to obtain nitrogen-doped porous carbon fibers. Accordingly, our approach is superior to the routine synthetic approaches for the preparation of nitrogen-doped porous carbons, which usually need acid washing and purification.

## 3. References to Supporting Information

## 4. Contributions

M.A. and J.Y. conceived the idea. J.G., M.A. and J.Y. planned and performed the experiments, collected and analyzed the data, and co-wrote the manuscript.